\documentclass[]{iopart}
\usepackage{iopams}
\usepackage{setstack}
\usepackage[]{graphicx}
\usepackage{color}
\usepackage[latin1]{inputenc}
\usepackage{marginnote}

\usepackage{mathbbol}
\usepackage{amssymb}             
\DeclareSymbolFontAlphabet{\amsmathbb}{AMSb}%

\newcommand{\tu}{\tilde{u}}

\begin{document}

\title[Stray light in the LISA's interferometer]{The LISA interferometer:
impact of stray light on the phase of the heterodyne signal}
\author{C P Sasso$^1$, G Mana$^1$, and S Mottini$^2$}
\address{$^1$INRIM -- Istituto Nazionale di Ricerca Metrologica, Str. delle cacce 91, 10135 Torino, Italy}
\address{$^2$Thales Alenia Space, Str. Antica di Collegno, 253, 10146 Torino, Italy}
\ead{c.sasso@inrim.it}

\begin{abstract}
The Laser Interferometer Space Antenna is a foreseen gravitational wave detector, which aims to detect $10^{-20}$ strains in the frequency range from 0.1 mHz to 1 Hz. It is a triangular constellation, with equal sides of $2.5 \times 10^9$ m, of three spacecrafts, where heterodyne interferometry measures the spacecraft distances. The stray light from the powerful transmitted beam can overlap with the received one and interfere with the heterodyne signal. We investigated the contribution of random phase variations of the stray photons to the noise of the heterodyne signal. A balanced detection scheme more effectively mitigates this adverse effect than a separation of the frequencies of the transmitted and local radiation. In the balanced scheme, in order to limit the phase noise to picometer level, the incoherent power of the stray light must be kept below about 10 nW/W for an asymmetry of the recombination beam splitter of 1$\%$.
\end{abstract}

\submitto{Classical and Quantum Gravity}


\section{Introduction}
The Laser Interferometer Space Antenna (LISA) is a concept for a space-based gravitational wave detector of the European Space Agency. It is a constellation of three spacecrafts -- an equilateral triangle with side length of $2.5 \times 10^6$ km -- trailing the Earth by 20 degrees. It aims to measure the fluctuations of the distance between free-falling masses placed inside the spacecrafts to picometre resolution in the frequency range from 0.1 mHz to 0.1 Hz. The detection of the test mass motions {\it vs.} the associated optical benches and the phase-linking of the local and remote benches by heterodyne interferometry split the measurement of the test-mass separation into onboard and inter-spacecraft interferometric measurements \cite{NGO:2011,Jennrich:2009,Weise:2017,Weise:2017-2}.

As regards the spacecraft separation, the phase of the weak received beam, about 200 pW, is detected by interfering it with a fraction of the transmitted beam, about 2 mW. Since both the beam transmission and reception are carried out by the same telescope and the same optical bench accommodates both the onboard and inter-spacecraft interferometers, the transmitted and received beams share part of their optical path \cite{d'Arcio:2017}. Consequently, although measures are taken to mitigate its detrimental effects, the light backscattered from the powerful transmitted beam, about 1 W, into the received one interferes with the measurement of the spacecraft distance and might jeopardise the sought picometer sensitivity \cite{Canuel:2013}.

Measures were proposed to mitigate the impact of stray light, such as tilted optical elements, baffling, off-axis telescopes, polarisation encoding interferometry \cite{Spector:2012,Livas:2015}. Also, measurement strategies, like the use of a balanced receiver \cite{Carleton:1968,Fleddermann:2018} and the swap of the local references between the two inter-spacecraft interferometers \cite{Weise:2017}, were investigated.

Experimental observations and an explaining model of the impact of back-reflected light on interferometric measurements are given in \cite{Hiroyuki:2000,Cavagnero:2005}. In previous papers, we investigated the phase noise of the inter-spacecraft interferometer due to the coupling of aberrated wavefronts with the transmitter and receiver jitters \cite{Sasso:2018a,Sasso:2018b}. In this paper, we estimate the contribution of back-reflected light.

In section \ref{s2}, we model the interference of stray light with the received and local beams and derive the measurement equations of the phase of the heterodyne signal for both the balanced signal-detection \cite{Carleton:1968,Fleddermann:2018} and the swap of the local references \cite{Weise:2017}. Stray light is not a problem per se, but its phase stability in the measurement bandwidth is. Hence, in section \ref{s3}, we quantify how the measurement error depends on the stray-light power, asymmetry of the recombination beam-splitter, and motion of the back-scattering elements.

\section{Stray light interference}\label{s2}
\subsection{Balanced detection}
The phase $\phi$ of the received wavefront is measured -- via optical heterodyne -- by interference with a fraction of the transmitted beam, which acts as a local reference. The beam transmission and reception are carried out by the same telescope, and the same bench accommodates both the transmission and reception optics. Therefore, the transmitted and received beams share some of the optical elements and some of the launched photons is sent back along the axis of the received beam, interfere with it, and contribute to the phase of the heterodyne signal. Since it senses the motion of the scattering elements, the backscattered light increases the measurement noise through its overall amplitude and phase stability in the measurement bandwidth.

A balanced detection of the heterodyne signal, where the beat notes in both output ports of the recombination beam splitter are detected and subtracted, can be applied to overcome this noise \cite{Carleton:1968,Fleddermann:2018}. In fact, since the signals of interest are in opposition while the parasitic ones are in phase, the subtraction rejects the noise stemming from the stray light.

\begin{figure}\centering
\includegraphics[width=7.5cm]{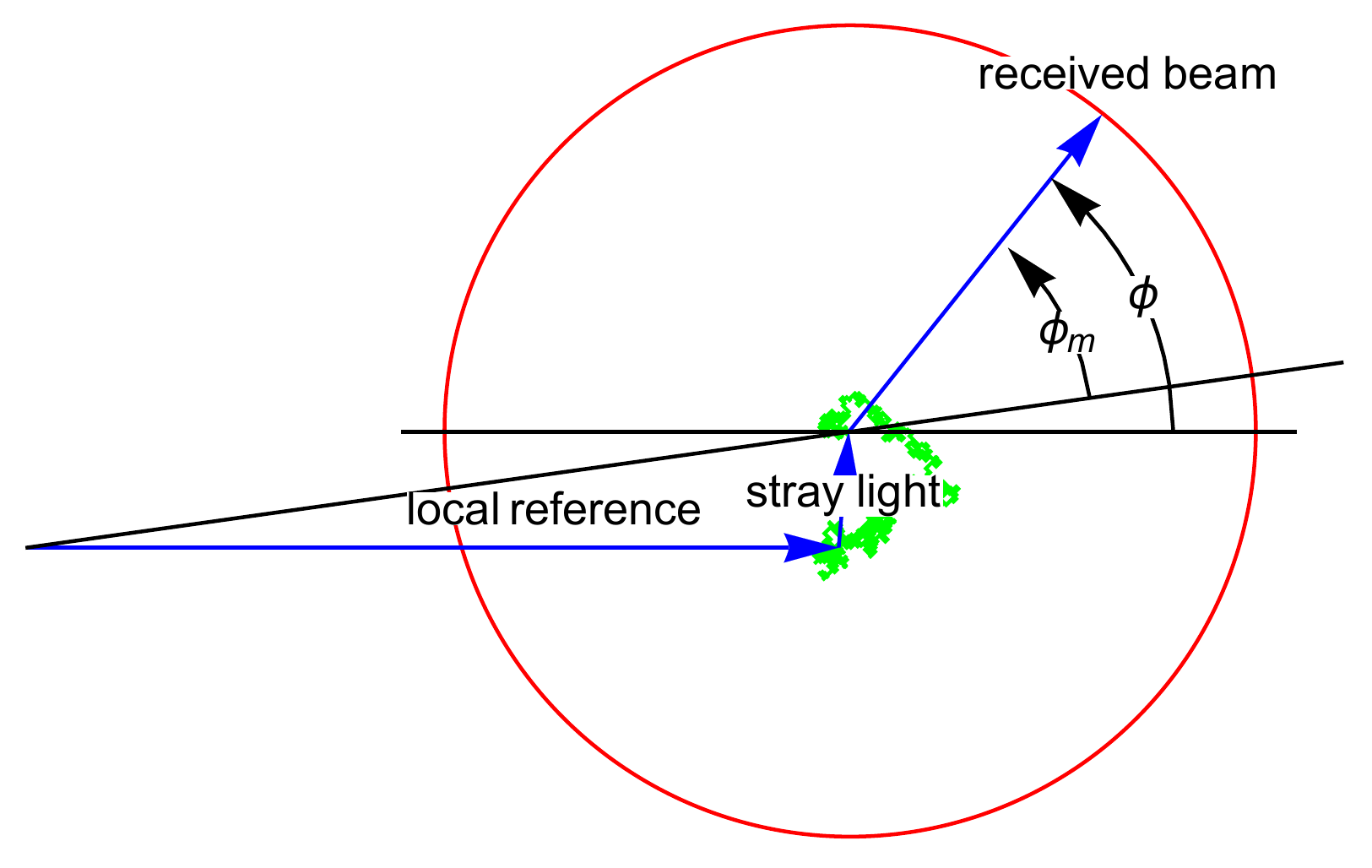}
\caption{Phasor diagram of the interfering optical fields. The light collected by the telescope rotates counterclockwise with angular velocity $\Omega$. The green random walk represents the stray light. The difference between $\phi_m$ and $\phi$ is the measurement error $\gamma$. When the power of the local beam tends to the infinity, the error tends to zero.} \label{Fig-01}
\end{figure}

As shown in Fig.\ \ref{Fig-01}, we model the stray light reaching the recombination beam splitter by the coherent sum of $N$ coaxial parasitic-rays,
\begin{equation}\label{sl}
 u\sqrt{2I_0}\rme^{\rmi(\omega t + \psi)} = \sqrt{2I_0} \sum_{n=1}^N u_n\rme^{\rmi \omega t+\psi_n} ,
\end{equation}
interfering with a fraction of the transmitted beam, $\sqrt{2I_0}\rme^{\rmi \omega t}$, and the light received from the remote spacecraft, $a\sqrt{2I_0}\rme^{\rmi(\omega+\Omega)t+\phi}$, where $2I_0 \approx 2$ mW is the power of the local reference, $a^2\approx \times 10^{-7}$ is the fractional power of the received beam, $\omega$ is the angular frequency of the transmitted and local beams, and $\phi$ is the phase of the received wavefront. Since the orbital dynamics causes a varying doppler shift, the heterodyne frequency $\Omega$ is continuously changing from 2 MHz to 19 MHz \cite{d'Arcio:2017}. In (\ref{sl}), $u_n^2$ and $\psi_n$ are the fractional power and phase of the parasitic optical fields, whereas $u^2$ and $\psi$ are those of their coherent sum.

The total field at output ports of the recombination beam splitters are
\numparts\begin{eqnarray}\label{many_beams}
 E_1 = 1 + a_1\rme^{\rmi(\Omega t+\phi+\pi/2)} +u_1\rme^{\rmi (\psi+\pi/2)} ,\\
 E_2 = \rme^{\rmi \pi/2} + a_2\rme^{\rmi(\Omega t+\phi)} +u_2\rme^{\rmi \psi} ,
\end{eqnarray}\endnumparts
where $i=1,2$ labels the output ports, $a^2 = a_1^2+a_2^2 \ll 1$, $u^2 = u_1^2+u_2^2 \ll 1$, and we omitted the common factor $\sqrt{I_0} \rme^{\rmi \omega t}$.  The heterodyne signals are
\numparts\begin{eqnarray}\fl
 S_1(t) = 1 + a_1^2 + u_1^2 - 2u_1\sin(\psi) - 2a_1\sin(\Omega t+\phi) + 2a_1u_1\cos(\Omega t+\phi-\psi) ,\\ \fl
 S_2(t) = 1 + a_2^2 + u_2^2 + 2u_2\sin(\psi) + 2a_2\sin(\Omega t+\phi) + 2a_2u_2\cos(\Omega t+\phi-\psi) .
\end{eqnarray}\endnumparts

The foreseen measurement of $\phi$ is based on a digital phase locked loop \cite{NGO:2011,Jennrich:2009,Shaddock:2006}. After the signal difference $S_2(t)-S_1(t)$ is anti-aliasing filtered, the digitised signal is multiplied with a local oscillator. The integrated output -- which is proportional to the phase difference between the signal and oscillator -- is used to lock the frequency of the local oscillator to $\Omega$. By modelling the phase recovery as
\numparts\begin{equation}
 \phi_m = \arg \left[ \int_0^{2n\pi/\Omega} [S_2(t)-S_1(t)] \rme^{-\rmi\Omega t}\, \rmd t \right] ,
\end{equation}
where the integration extends over $n$ oscillator cycles, the measurement equation of the heterodyne-signal phase is
\begin{equation}\label{mphase}
 \phi_m \approx \phi + \tu\sin(\psi) + \pi/2 ,
\end{equation}\endnumparts
where, since the heterodyne signals are balanced so as to have the same alternating amplitude, $a_1\approx a_2$ and
\begin{equation}
 \tu = \frac{a_1u_1-a_2u_2}{a_1+a_2} \approx \frac{u_1-u_2}{2} .
\end{equation}
For later convenience, we redefined $\psi$ as $\psi-\pi/2$, assumed $u_{1,2} \ll 1$, and considered only the terms up to the first order.

If the recombination beam splitter deviates from a 50:50 power-splitting ratio, after adjusting the alternating components of the heterodyne signal in such a way that $a_1 = a_2 = a/\sqrt{2}$, the total coherent-amplitudes of the stray light at the output ports of the interferometers are
\begin{equation}
 u_{1,2} = \sqrt{\frac{1 \pm \epsilon}{2}}\, u \approx \frac{(1 \pm \epsilon/2)u}{\sqrt{2}}
\end{equation}
where $(1 + \epsilon)/(1 - \epsilon)$ is the ratio of the reflected to the transmitted powers. Hence, the measurement error $\phi_m-\phi$ in (\ref{mphase}) is
\begin{equation}\label{pe}
 \gamma = \frac{\epsilon u\sin(\psi)}{2\sqrt{2}} .
\end{equation}
The non-balanced detection can be modelled by letting $a_1$ or $a_2$ go to zero. In this case,
\begin{equation}
 \gamma = \pm u\sin(\psi) .
\end{equation}

\subsection{Frequency swap}
A different proposal to overcome the stray-light issue is to introduce an offset between the frequencies of the local and transmitted beams \cite{Weise:2017}. This shift is obtained by swapping the local references between the two spacecraft's optical benches and by operating the two laser sources at different frequencies.

The total field at the interferometer detector is
\begin{equation}
 E = 1 + a\rme^{\rmi(\Omega_1 t+\phi)} + u\rme^{\rmi(\Omega_2 t+\psi)} ,
\end{equation}
were $\Omega_{1,2}$ are the frequency offsets ({\it vs}.\ the local reference) of the received and transmitted beams, respectively. For the sake of simplicity we did not consider a balanced detection and omitted again the $\sqrt{I_0} \rme^{\rmi \omega t}$ factor. Hence, the heterodyne signal is
\begin{equation}\fl
 S(t) = 1 + a^2 + u^2 + 2u\cos(\Omega_2 t+\psi) + 2a\cos(\Omega_1 t+\phi) + 2au\cos(\Delta_\Omega t+\phi-\psi) ,
\end{equation}
where $\Delta_\Omega = \Omega_1-\Omega_2$.

\begin{figure}\centering
\includegraphics[width=6.25cm]{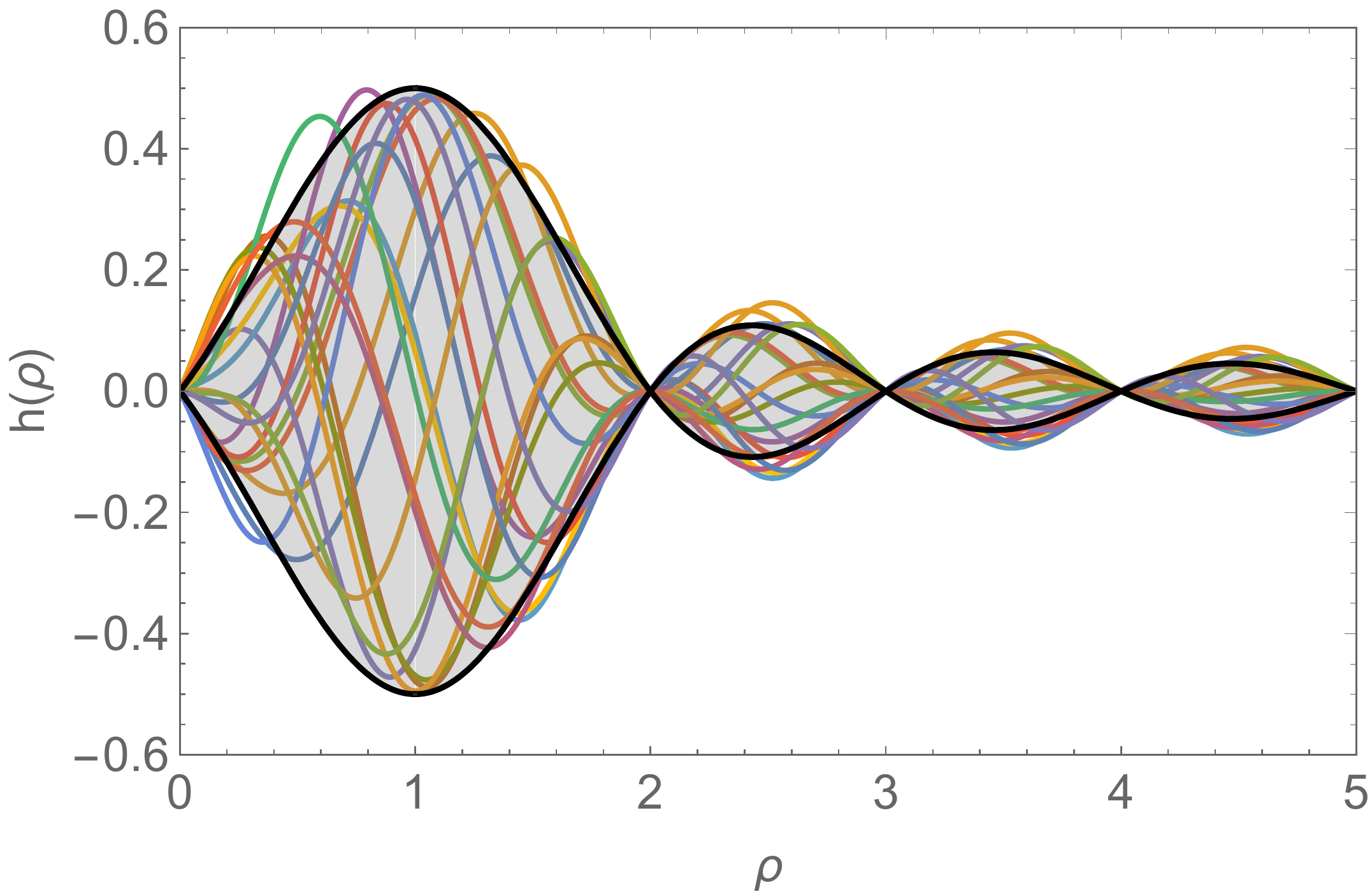}
\includegraphics[width=6.25cm]{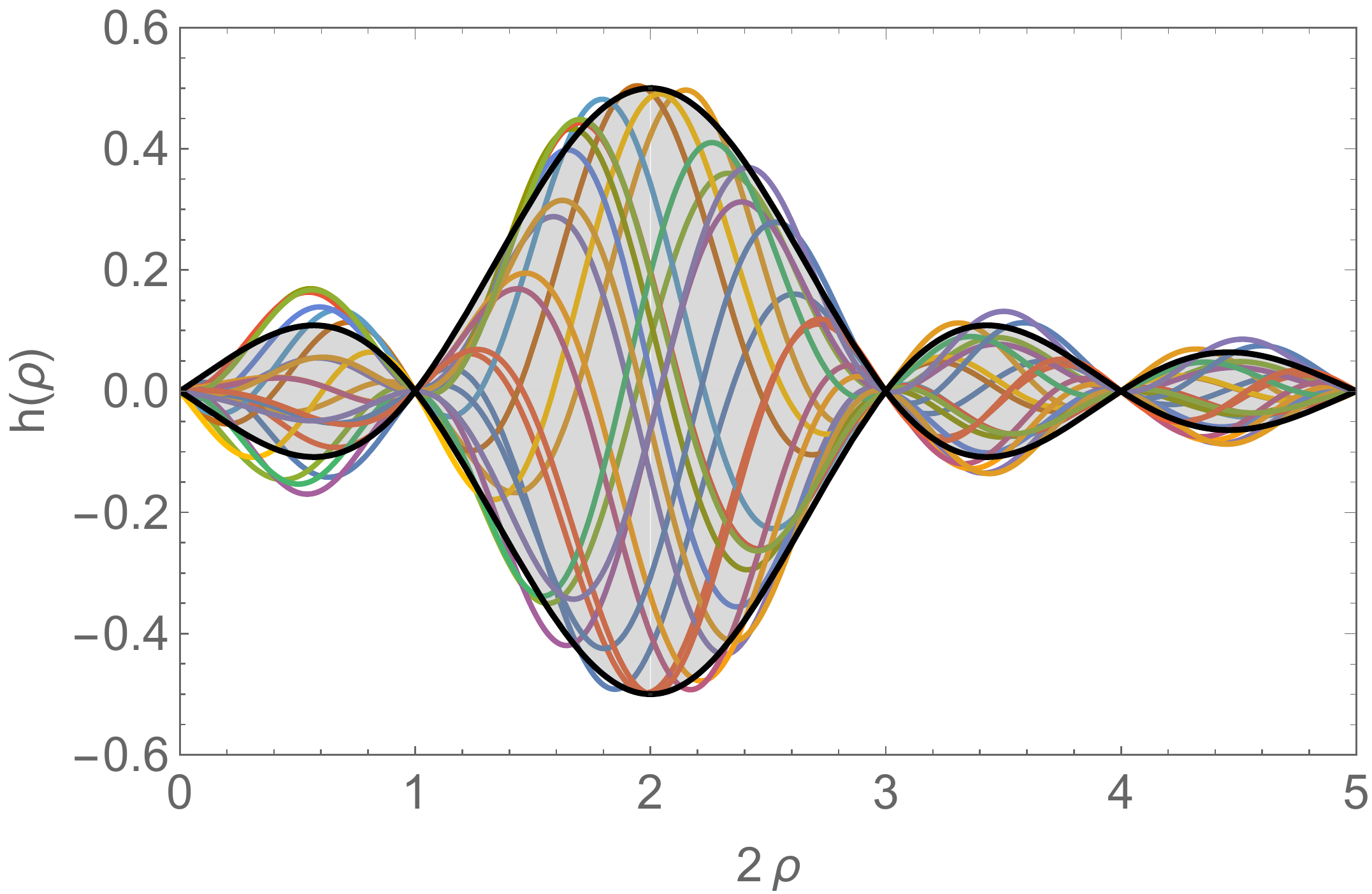}
\caption{$h(\rho)$ when $n=$ 1 (left) and 2 (right). The coloured lines correspond to random values of the received wavefront and stray-light phases. The gray area is the approximate envelope. The fractional power of the local beam is $a^2=10^{-7}$.} \label{Fig-swap1}
\end{figure}

By assuming $au \ll u \ll a \ll 1$, the frequency of the local oscillator locks to $\Omega_1$ for all practical purposes. Therefore, the phase measurement-equation is
\begin{equation}\label{swap-error}
 \phi_m = \arg \left[ \int_0^{2n\pi/\Omega_1} S(t) \rme^{-\rmi\Omega_1 t}\, \rmd t \right]
 \approx \phi + \frac{2u h(\rho)}{a} ,
\end{equation}
where
\begin{eqnarray}\nonumber\fl
  h(\rho) = &\frac{\sin(n\pi\rho)}{n\pi\rho}
                 \bigg[ \frac{\rho^2\cos(n\pi\rho+\psi)\sin(\phi)}{1-\rho^2} - \frac{a(1-\rho)\cos(n\pi\rho+\psi-\phi)\sin(\phi)}{2-\rho}
  \\ \fl \label{hrho}
                &+\frac{\rho\sin(n\pi\rho+\psi)\cos(\phi)}{1-\rho^2} - \frac{a\sin(n\pi\rho+\psi-\phi)\cos(\phi)}{2-\rho} \bigg] ,
\end{eqnarray}
$\rho=\Omega_2/\Omega_1$ and the needed integration were carried out with the aid of Mathematica$^{\circledR}$ \cite{Mathematica}. The zeroes of $h(\rho)$ occurs at the integer $\rho$ values (both positive and negative), but
\begin{equation}\label{zero}
  h(\rho=0) = -a\sin(\psi)/2 ,
\end{equation}
which corresponds to the no frequency swap. By neglecting the terms proportional to $a$, as shown in Fig. \ref{Fig-swap1}, the approximate envelope of (\ref{hrho}) is
\begin{equation}\label{envelope}
 |h(\rho)| \lesssim \frac{|\sin(n\pi\rho)|}{2\pi n|1-\rho|} .
\end{equation}
Since the heterodyne frequency $\Omega_1$ and, consequently, the phase $\phi$ vary continuously, we can bound the measurement error $\gamma=\phi_m-\phi$ in (\ref{swap-error}) by
\begin{equation}\label{env2}
 |\gamma| \lesssim \frac{u|\sin(n\pi\rho)|}{\pi a n|1-\rho|} .
\end{equation}
It is worth to note that, in order to ensure noise rejection, it is necessary to satisfy the constraint $|n\rho| \gg 1$.

\section{Phase noise}\label{s3}
\subsection{Balanced detection}
The phases of the parasitic rays $u_n\rme^{\rmi\psi_n}$ in (\ref{sl}) are unpredictable. Therefore, in order to evaluate their effect on the heterodyne signal, let us assume that the motion of every scattering element is uniformly distributed in the $[-\lambda/2, +\lambda/2]$ interval, where $\lambda = 1064$ nm is the wavelength. A less extremal assumption will be discussed later. Hence,
\begin{equation}
 \sin(\psi_n) \sim \frac{1}{\pi\sqrt{1-z_n^2}} ,
\end{equation}
where the tilde means "is distributed as" and $-1 < z_n <1$ are the $\sin(\psi_n)$ values.  Remembering that
\begin{equation}\fl
 u\cos(\psi) + \rmi u\sin(\psi) =
 u\rme^{\rmi\psi} = \sum_{n=1}^N u_n \cos(\psi_n) + \rmi\sum_{n=1}^N u_n \sin(\psi_n) ,
\end{equation}
we obtain
\begin{equation}
 \langle u\sin(\psi)\rangle = \left\langle \sum_{n=1}^N u_n \sin(\psi_n) \right\rangle = 0
\end{equation}
and, since ${\rm var}[\sin(\psi_n)] = 1/2$,
\begin{equation}\label{var-us}
 {\rm var}[u\sin(\psi)] = \frac{1}{2} \sum_{n=1}^N u_n^2 = \frac{I_{\rm i}}{2I_0} ,
\end{equation}
where $2I_{\rm i} = 2I_0 \sum_{n=1}^N u_n^2$ is the total incoherent power of the stray light and $2I_0$ is the power of the local beam. Therefore, on the average, the phase error (\ref{pe}) is null. Its variance is
\begin{equation}\label{sdv}
 \sigma_\gamma^2 = \frac{\epsilon I_{\rm i}}{4\sqrt{2}\,I_0} .
\end{equation}

To give a numerical example, the requirement $\sigma_\gamma \lambda/(2\pi) < 1$ pm -- or, being $\lambda = 1064$ nm, $\sigma_\gamma /(2\pi) < 10^{-6}$ -- constrains the total incoherent-power of the stray light to
\begin{equation}\label{c1}
 I_{\rm i} \lesssim \frac{2\times 10^{-10} I_0}{\epsilon} .
\end{equation}

Given the otherwise required high stability of the optical assembly, to examine the impact of realistic elements' motions, we assume that the phases $\psi_n$ in (\ref{sl}) are uniformly walking in the intervals $[\psi_{n0}-\alpha, \psi_{n0}+\alpha]$, where $\alpha \ll 1$ rad. Since now
\begin{equation}
 {\rm var}[u_n\sin(\psi_n)] \approx \frac{\alpha^2 u_n^2 \cos^2(\psi_{0n})}{3} ,
\end{equation}
(\ref{var-us}) must be updated to
\begin{equation}
  \begin{array}{ll}
    {\rm var}[u\sin(\psi)] &= \displaystyle \frac{\alpha^2 \sum_{n=1}^N u_n^2 \cos^2(\psi_{0n})}{3}  \\
                           &\approx \displaystyle \frac{\alpha^2 \sum_{n=1}^N u_n^2}{6} = \frac{\alpha^2 I_{\rm i}}{6I_0}
  \end{array} ,
\end{equation}
where we assumed $\psi_{0n}$ uniformly distributed in the $[-\pi, \pi]$ interval and substituted the $1/2$ average for $\cos^2(\psi_{0n})$. Therefore, the variance of the phase error (\ref{pe}) is
\begin{equation}\label{s22}
 \sigma_\gamma^2 = \frac{\epsilon \alpha^2 I_{\rm i}}{12\sqrt{2}\,I_0} .
\end{equation}
Consequently, the constrain (\ref{c1}) relaxes to
\begin{equation}\label{c2}
 I_{\rm i} \lesssim \frac{6\times 10^{-10} I_0}{\epsilon\alpha^2} .
\end{equation}

\subsection{Frequency swap}
The continuous change of the heterodyne frequency $\Omega_1$ and phase $\phi$  due to the spacecraft's orbit-dynamics makes the phase error in (\ref{swap-error}) varying also if the scattering elements do not move. By assuming the phase error uniformly distributed within the bounds (\ref{env2}), the noise variance is bounded by
\begin{equation}\label{s23}
 \sigma^2_\gamma \lesssim \frac{1}{3} \left[ \frac{u\sin(n\pi\rho)}{\pi a n(1-\rho)} \right]^2
 \lesssim \frac{I_{\rm c}}{3I_1} \left[ \frac{1}{\pi n(1-\rho)} \right]^2  ,
\end{equation}
where $I_c$ is the total coherent-power of the stray light, $I_1$ is the received power, and we increased $\sin^2(n\pi\rho)$ to one.

To give again a numerical example, the sought $\sigma_\gamma \lambda/(2\pi) < 1$ pm target requires that total coherent-power of the stray light is constrained by
\begin{equation}
 I_{\rm c} \lesssim 1.2 \times 10^{-9} n^2 (1-\rho)^2 I_1 \approx 1.2 \times 10^{-16} n^2 (1-\rho)^2 I_0 ,
\end{equation}
where we used $I_1=10^{-7} I_0$. Therefore, to make (\ref{s23}) competitive against (\ref{sdv}) and (\ref{s22}), a huge number $n$ of integration cycles is necessary.

\section{Conclusions}
The separation of the LISA's spacecraft is monitored by heterodyne interferometry down to picometre sensitivity, in which laser beams are transmitted and received by the same telescopes. The phase of the received wavefront is detected by mixing it with a fraction of the transmitted one. The light backscattered from the transmitted beam into the received one has a detrimental effect on the interferometric measurement and, therefore, it is a critical issue. As long as it has a fixed phase, it does not limit the measurement, but any phase variations cause a noise.

We reported noise estimates for the balanced detection of the heterodyne signal and the swap of the local references and quantified how the noise depends on the stray-light power, asymmetry of the recombination beam-splitter, and motion of the back-scattering elements. Since the spacecraft's orbit-dynamics prevent optimisation, the swap of the local references is not appealing. By using the equation (\ref{c1}), in order to limit the phase noise to picometer level, the incoherent power of the stray light must be kept below about 10 nW/W for an asymmetry of the recombination beam splitter of 1$\%$. Our results are of importance to the interferometer design and underpin the specifications for the manufacturing of the inter-spacecraft interferometer and the estimate of the error budget.

\section{Acknowledgments}
This work was funded by the European Space Agency (contract 1550005721, Metrology Telescope Design for a Gravitational Wave Observatory Mission).

\section*{References}
\bibliography{LISA}

\end{document}